\documentclass[acmtog]{acmart}
\pdfoutput=1

\usepackage{hyperref}
\usepackage{booktabs} 
\usepackage{microtype}
\usepackage{amsmath}
\usepackage{multirow}
\usepackage{multicol}
\usepackage{makecell}
\usepackage{diagbox,pict2e}
\newcommand{\STAB}[1]{\begin{tabular}{@{}c@{}}#1\end{tabular}}
\DeclareMathOperator*{\argmin}{arg\,min}

\setcitestyle{square}

\author{Meng Zhang}
\affiliation{ \institution{University College London} } 

\author{Tuanfeng Wang}
\affiliation{ \institution{miHoYo Inc.} } 

\author{Duygu Ceylan}
\affiliation{ \institution{Adobe Research} } 

\author{Niloy J. Mitra}
\affiliation{ \institution{University College London and Adobe Research} }

\renewcommand\footnotetextcopyrightpermission[1]{}
\begin{document}

\title{Deep Detail Enhancement for Any Garment}

\fancyfoot{}
\begin{abstract}
Creating fine garment details requires significant efforts and huge computational resources. In contrast, a coarse shape may be easy to acquire in many scenarios (e.g., via low-resolution physically-based simulation, linear blend skinning driven by skeletal motion, portable scanners). In this paper, we show how to enhance, in a data-driven manner, rich yet plausible details starting from a coarse garment geometry. Once the parameterization of the garment is given, we formulate the task as a style transfer problem over the space of associated normal maps. In order to facilitate generalization across garment types and character motions, we introduce a patch-based formulation, that produces high-resolution details by matching a Gram matrix based style loss, to hallucinate geometric details (i.e., wrinkle density and shape). We extensively evaluate our method on a variety of production scenarios and show that our method is simple, light-weight, efficient, and generalizes across underlying 
garment types, sewing patterns, and body motion. 
\end{abstract}



\begin{teaserfigure}
  \centering
  \includegraphics[width=\textwidth]{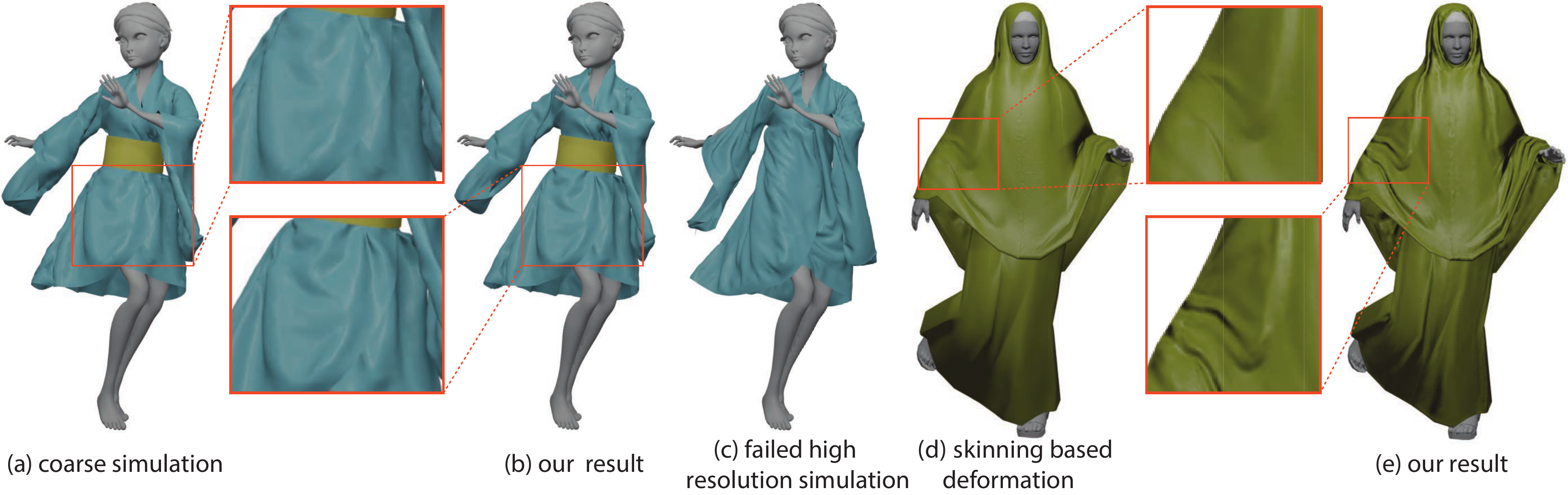}
\caption{
We present a data-driven approach to synthesize plausible wrinkle details on a coarse garment geometry. Such coarse geometry can be obtained either by running a physically-based simulation on a low resolution version of the garment~(a), or by using a skinning-based deformation method~(d). Our method can generate fine-scale geometry~(b,e) by replacing the expensive simulation process, which, in some cases, are not even currently feasible to setup. For example, high-resolution simulation  failed~(c) for the blue dress as boundary conditions between the blue dress and the yellow belt were grossly violated. Ours generalizes across garment types, often with different number of patches, associated parameterization and materials, and undergoing different body motions.
}
\label{fig:teaser}
\end{teaserfigure}

\maketitle

\thispagestyle{empty}

\section{Introduction}
High fidelity garments are a vital component of many applications including virtual try-on and realistic 3D character design for games, VR, and movies. In real life, garments, as they undergo motion, develop folds and rich details, characteristic of the underlying materials. Naturally, we desire to carry over such realistic details to the digital world as well. 

\begin{figure}[t]
  \includegraphics[width=\columnwidth]{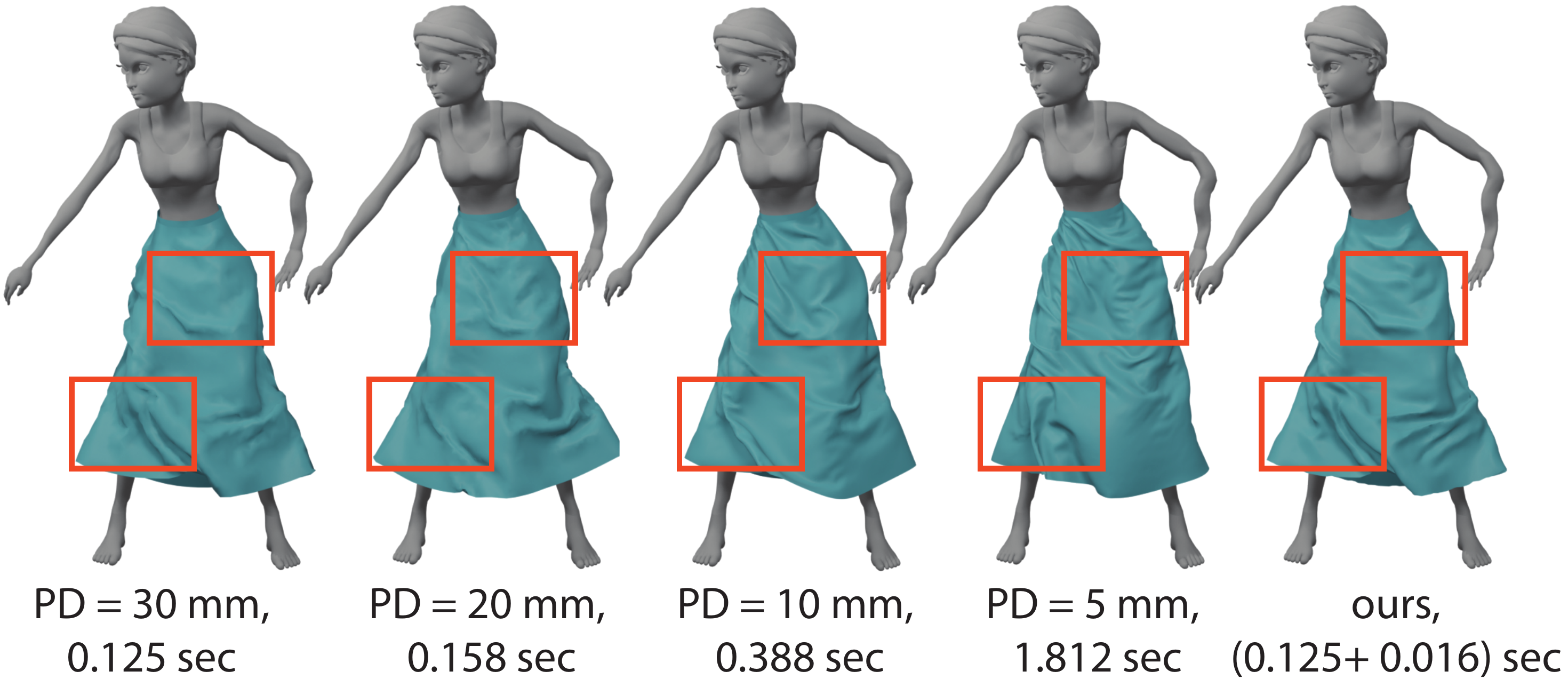}
  \caption{We show a garment simulated with different mesh resolutions, i.e., varying particle distances (PD), and report the computation time for simulating a single frame. While wrinkle details are captured well when simulating high resolution meshes, i.e., lower PD, computation time significantly increases. Our method captures statistically similar quality details by enhancing a coarse simulation much more efficiently ($13$x speedup here).}
  \label{fig:motivation}
\end{figure}

Directly capturing such fine geometric details requires professional capture setups along with  computationally-heavy processing pipeline~\cite{bradley2008markerless,chen2015garment,pons2017clothcap}, which may not be applicable for low-budget scenarios. An alternate option involves physically-based cloth simulation~\cite{narain2012adaptive,tang2018cloth,liang2019differentiable} to virtually replace the need for physical setups. However, despite many research advances, achieving high quality and robust simulation still has many challenges. The simulation is often sensitive to the choice of parameters and initial conditions and may require a considerable amount of manual work to setup correctly. Even when setup, it remains a computationally expensive process, especially as the resolution of the garment geometry increases (see Figure~\ref{fig:motivation}). Further, with  increasing complexity of the garment geometry, e.g., garments with many layers or with non-canonical UV layout, it may be difficult to obtain any stable result (see Figure~\ref{fig:teaser}(c)). Finally, for each new garment type and/or motion sequences, majority of the setup and simulation steps have to be repeated.

We propose a different approach. 
As garment undergoes motion, its geometry can be analyzed in two steps. First, the corresponding motion of the body (i.e., the body pose) changes the global shape of the garment, e.g., its boundary and coarse folds. We observe that the global shape to be consistent across different resolutions of a garment (see Figure~\ref{fig:motivation}). Such a global geometry can either be simulated efficiently and reliably using a coarsened version of the garment,  or by utilizing a skinning based deformation method. 
Given the coarse geometry, the high frequency wrinkles can be perceived by considering only local neighborhood information. As shown in Figure~\ref{fig:motivation}, the highlighted local patches in the coarse garment geometry provide sufficient cues to hallucinate plausible high frequency wrinkles in the same region. Hence, we hypothesize that given the coarse geometry of a garment, the dynamics of the detailed wrinkles can be learned from many such local patches observed from different type of garments undergoing different motions.

In this work, we present a deep learning based method that enhances an input coarse garment geometry by synthesizing realistic and plausible wrinkle details. Inspired by the recent success of image-based artistic style transfer methods~\cite{gatys2016image}, we represent the garment geometry as a 2D normal map where we interpret the coarse normal map as the \emph{content} and fine geometric details as a \emph{style} that can be well captured by the correlation between VGG features captured in the form of Gram matrices, characteristic to different materials. Furthermore, in order to tackle various garment materials within a universal framework, we adopt the conditional instance normalization technique~\cite{dumoulin2016learned} that is proposed in the context of universal style transfer. Finally, we design our detail enhancement network to operate at a patch level enabling generalization across different (i)~garment types, (ii)~sewing patterns, and (iii)~underlying motion sequences. In fact, we show that our network trained on a dataset generated from only a single sheet of cloth by applying random motion sequences (e.g., applying wind force, being hit by a ball) can already reasonably generalize to various garment types such as a skirt, a shirt, or a dress. See Figure~\ref{fig:ablation} for a comparison. 

We evaluate our approach on a challenging set of examples demonstrating generalization both across complex garment types (e.g., with layering) as well as a wide range of motion sequences. In addition to enriching low resolution garment simulations with accurate wrinkles, we also demonstrate that our method can enhance coarse garment geometries obtained by skinning methods as often used by computer games, see Figure~\ref{fig:teaser}. To our knowledge, we present the first learning-based method for generating garment details that generalizes to unseen garment types, sewing patterns, and body motions all at the same time.

\section{Related Work}

Realistic and physically based cloth simulation has been extensively studied in the graphics community~\cite{Ko2005,Nealen2006,narain2012adaptive,liang2019differentiable,yu2019simulcap,tang2018cloth}. However, due to the associated computational cost and stability concerns, as the complexity of the garments increase, several alternative avenues have been explored. For example, in the industrial settings, skinning based deformation of garments is often favored for its simplicity, robustness, and efficiency. This, however, comes at the expense of losing geometric details. Finally, there has been various efforts to augment coarse simulation output with fine details using various approaches including optimization, data-driven, and learning-based methods. 

\paragraph{Optimization-based methods.}
M\"{u}ller et al.~\shortcite{Muller2010} use a constraint-based optimization method to compute a high resolution mesh from a base coarse mesh.  Rohmer et al.~\shortcite{Rohmer2010} uses  stretch tensor of the coarse garment animation as a guide to place wrinkles in a post-processing stage. However, they assume the underlying garment motion is smooth which is not always the case with character body motion. Gillette et al.~\shortcite{Gillette2015} extends similar ideas to real time animation of cloth.

\paragraph{Data-driven methods.}  
Feng et al.~\shortcite{Feng2010} present a deformation transfer system that learns a non-linear mapping from a coarse garment shape to per-vertex displacements that represent fine details. Zurdo et al.~\shortcite{zurdo2012animating} share a similar idea to learn the mapping from a low-resolution simulation of a cloth to high resolution where plausible wrinkles are defined as displacements. Given a set of coarse and mid-scale mesh pairs, Kavan et al.~\shortcite{Kavan2011} learn linear upsampling operators. Wang et al.~\shortcite{Wang2010} construct a database of high resolution simulation examples generated by moving each body joint separately. Given a coarse mesh, wrinkle regions for each joint are synthesized by interpolating between database samples and blended into a final mesh. However, this method is limited to tight fitting clothing. One of the most comprehensive system in this direction is the DRAPE~\cite{guan2012drape} system that learns a linear subspace model of garment deformation driven by pose variation. Hahn et al.~\shortcite{hahn2014subspace} extend this idea by using adaptive bases to generate richer deformations. Xu et al.~\shortcite{xu2014sensitivity} introduce a pose-dependent rigging solution to extract nearby examples of a cloth and blend them in a sensitive-based scheme to generate the final drape.

\paragraph{Learning-based methods.}
With the recent success of deep learning methods in various imaging and 3D geometry tasks, a recent research direction is to \emph{learn} garment deformations under body motion using neural networks. A popular approach has been to extend parametric human body models with a per-vertex displacement map to capture the deforming garment geometry ~\cite{alldieck2019learning,bhatnagar2019multi,pons2017clothcap,jin2018pixel}. While this is a very efficient representation, it only works well for tight clothes such as t-shirt or pants that are sufficiently close to the body surface. Gundogdu et al.~\shortcite{gundogdu2019garnet} present GarNet which learns features of garment deformation as a function of body pose and shape at different levels (point-wise, patch-wise, and global features)  to reconstruct the detailed draping output. Santesteban et al.~\shortcite{santesteban2019learning} learn a garment deformation model via recurrent neural networks that enables real-time virtual try-on. Their approach learns the coarse garment shape based on the body shape and the detailed wrinkles based on the pose dynamics. A similar two-stage strategy is proposed by Patel et al.~\shortcite{patel2020virtual} which decomposes the deformation caused by body pose, shape, and garment style, into high frequency and low-frequency components. Wang et al.~\shortcite{wang2019learning} also consider garment deformations due to material properties in their motion-driven network architecture. Others~\cite{yang2018analyzing, wang2018learning, holden2019subspace} investigate a subspace technique such as principal component analysis, to reduce the number of variables of the high dimensional garment shape space. While these learning-based methods are efficient and fully differentiable compared to standard physically-based simulation, they are hard to generalize across different garment types and associated sewing patterns. In this work, we seek to combine the best of both worlds,  physically-based simulation and learning-based approach. Specifically, we rely on physically-based simulation for fast coarse garment generation, we synthesize fine geometry conditioned on the coarse shape by a neural network. By utilizing a patch-based approach, and training on synthetic data, we demonstrate that our trained model generalizes across different garment types as well as different material configurations, and also extends to coarse geometry obtained using linear blend skin.

One of the closest learning based methods to our approach is DeepWrinkles~\cite{lahner2018deepwrinkles} which augments a low-resolution normal map of a garment captured by RGB-D sensors with high-frequency details using PatchGAN. However, their model is garment-specific and falls short when applied to garments unseen during training (e.g., garments with a different uv-parameterization or material properties).  In Section~\ref{sec:evaluation}, we show a comparison of DeepWrinkle to contrast against ours generalization ability, wherein we use conditional instance normalization to further adapt to  material specifications and directly match Gram matrix based features, instead of matching style using a GAN. 

Generalizing learning based methods to different garment types with potentially different number of vertices, connectivity, and material properties is an important challenge. In order to address this challenge, the recent work of Ma et al. \shortcite{ma2019learning} adopts a one-hot vector to encode the \textit{pre-defined} garment type and trains a graph convolutional neural network with a generative framework to deform the body mesh into the garment surface. Zhu et al.~\shortcite{zhu2020deep} train an adaptable template for image based garment reconstruction and introduce a large garment dataset with more than $2K$ real 3D garment scans spanning over 10 different categories. This method generates plausible results if an initial template with a reasonable topology is provided. Alldieck et al. ~\shortcite{alldieck2019tex2shape} use normal and displacement maps based on a fixed body UV parameterization to represent the surface geometry of static garments. The fixed parameterization assumption makes it difficult to generalize the method to new garment designs, especially more complex garments such as multi-layer dresses. We also utilize a normal map representation, however, our patch-based method focuses on local shape variations and is independent of the underlying UV parameterization.

\paragraph{Image-to-image transfer.}
We draw inspiration from the recent advances in image-to-image transfer methods~\cite{liao2017visual,fivser2016stylit,huang2017arbitrary,isola2017image} as we represent the garment geometry as a normal map and cast the problem as detail synthesis in this 2D domain. Specifically, we follow the framework proposed by~\cite{gatys2015texture,gatys2016image} that captures the style of an image by the Gram matrices of its VGG features~\cite{simonyan2014very}. Similarly, we assume that a coarse normal map can be enhanced with fine scale wrinkles by matching the feature statistics of Gram matrices. Our work can be loosely related to image  super-resolution where, the aim is to synthesize high-resolution images from blurry counterparts~\cite{ledig2017photo,wang2015deep}.
In Section~\ref{sec:evaluation}, we provide comparison with a direct image space superresolution approach. 


\begin{figure*}[t!]
  \includegraphics[width=\textwidth]{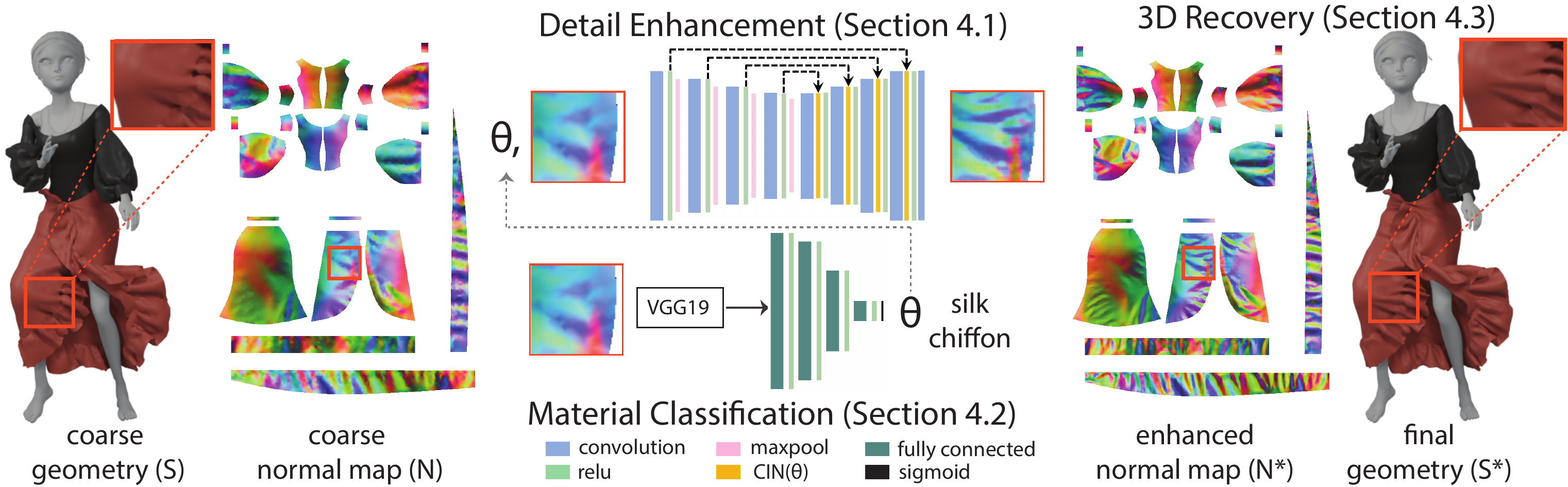}
  \caption{Given a coarse garment geometry $\mathbb{S}$ represented as a 2D normal map $\mathbb{N}$, we present a method to generate an enhanced normal map $\mathbb{N^*}$ and the corresponding high resolution geometry $\mathbb{S^*}$. At the core of our method is a detail enhancement network that enhances local garment patches in $\mathbb{N}$ conditioned on the material type $\theta$ of the garment. We combine such enhanced local patches to generate $\mathbb{N^*}$ which captures fine wrinkle details. We lift $\mathbb{N^*}$ to 3D to generate $\mathbb{S^*}$ by an optimization method that avoids interpenetrations between the garment and the underlying body surface. In case the garment material is not known in advance, we present a material classification network that operates on the local patches cropped from the coarse normal map $\mathbb{N}$.}
  \label{fig:overview}
\end{figure*}

\section{Overview}
Our goal is to augment a coarse garment geometry source sequence, $\mathbb{S}$, with fine details such as plausible and realistic wrinkles. We assume that $\mathbb{S}$ is obtained by either a physically-based simulation run on a low-resolution version of the garment or obtained by computational fast approaches such as skinning based deformation. The output of our pipeline is a fine-detailed surface sequence, $\mathbb{S^*}$, which depicts the source garment with realistic wrinkles in a temporally-consistent manner. The fine wrinkles in $\mathbb{S^*}$ are synthesized either by \textit{sharpening} the original coarse folds in $\mathbb{S}$ or \textit{synthesized} conditioned on nearby coarse folds. 

Motivated by the recent advances in image style transfer~\cite{gatys2016image}, we cast the 3D wrinkle synthesis task as a \emph{2D detail transfer} problem defined on the normal maps of the garments. Specifically, we present a learning-based method to learn an image-to-image transfer network $\Phi$ that synthesizes a detail-rich normal map $\mathbb{N^*}$ corresponding to $\mathbb{S^*}$ given the coarse normal map $\mathbb{N}$ of $\mathbb{S}$ by matching the feature statistics using Gram Matrices (Section~\ref{subsec: LearningForStyleTransfer}). 

However, unlike previous work~\cite{lahner2018deepwrinkles}, which focuses on learning garment-specific models using PatchGAN, our goal is to train a \textit{universal detail transfer network} that can generalize over different types of garments (e.g., dress, shirt, skirt) made of different materials (e.g., silk, leather, matte) and undergoing different body motions (e.g., dancing, physical activities). First, in order to ensure generalization across different materials and body motions, we adopt the conditional instance normalization (CIN) approach, introduced by Dumoulin et al.~\shortcite{dumoulin2016learned} in the context of artistic style transfer. Specifically, given the material type $\theta$ of a garment, we shift and scale the learned network parameters to adapt to different materials and motions. Given a sequence of coarse garment geometry, we predict its material type $\theta$ via a classifier, and do not require the material specification as part of the input. 
Second, in order to generalize across different types of garments and different uv-parameterizations utilized to generate the normal maps of the garments, we propose a novel patch-based approach. Specifically, our network works at a patch level by synthesizing details on overlapping patches randomly cropped from the normal map, $\mathbb{N}$, of the garment, $\mathbb{S}$. We merge the resulting detailed patches to obtain the final normal map, $\mathbb{N^*}$, which is lifted to 3D via a gradient-based optimization scheme to yield $\mathbb{S^*}$ (Section~\ref{subsec:3dlifting}). Our pipeline is illustrated in Figure~\ref{fig:overview}.

\section{Algorithm Stages}

Given a coarse 3D garment geometry sequence $\mathbb{S}$, for each frame $\mathbb{S}_i \in \mathbb{S}$, our goal is to synthesize $\mathbb{S}_i^*$, which enhances the garment geometry with fine wrinkle details, subject to the underlying body motion and the assigned garment material $\theta$ (see Section~\ref{subsec:garmentClassify}). Note that given the nature of garments, any $\mathbb{S}_{i}$ can be semantically cut into a few developable pieces, i.e., garment patterns. We assume such a parameterization is provided for the first frame $\mathbb{S}_0$ and applicable to the remaining frames in the sequence.

\subsection{Detail Enhancement}\label{subsec: LearningForStyleTransfer}
Using the garment-specific parameterization, we represent each $\mathbb{S}_i$ by a normal map $\mathbb{N}_i$. We treat the detail enhancement problem as generating a normal map $\mathbb{N}_i^*$ conditioned on $\mathbb{N}_i$ and the material characteristics $\theta$. While $\mathbb{N}_i^*$ preserves the content of $\mathbb{N}_i$ (e.g., the density and the position of the wrinkles), the appearance of the details (e.g., shape, granularity) are learned from $\{\mathbb{N}_{\theta}\}$, a collection of normal maps depicting the detailed geometry of various garments of material $\theta$ undergoing different body motions. The detail enhancement step operates at each frame, hence, for simplicity, we ignore the subscript $i$  in the following.

\paragraph{Detail Enhancement Network.}
We adopt a U-net architecture for our detail enhancement network (see Figure~\ref{fig:overview}). The encoder projects the input $\mathbb{N}$ into a learned latent space. The latent representation $\mathbf{z}$ is then used to construct the output $\mathbb{N}^*$ via the decoder. Although garments made of different materials may share similar coarse geometry, the fine details can vary, conditioned on the material. To adapt our approach to different garment materials and capture fine details, we use $\theta$, a one-hot vector, as input to indicate the type of material $\mathbb{N}$ and $\mathbb{N}^*$ are associated with. We design the decoder using conditional instance normalization (CIN)~\cite{dumoulin2016learned} to adapt for different material types. Specifically, given a material type $\theta$, the activation of a layer $i$ is converted to a normalized activation $\mathbf{z}_{i}^\theta$ specific to a material type $\theta$ according to, 
\begin{equation*}
\mathbf{z}_{i}^\theta := \gamma_{i}^\theta \left(\frac{\mathbf{x}_{i}-\mu(\mathbf{x}_{i})}{\sigma(\mathbf{x}_{i})}\right) + \epsilon_{i}^\theta,
\end{equation*}
where $\mu$ and $\sigma$ are the mean and standard deviation of $\mathbf{x}_{i}$, and $\gamma_{i}^\theta$ and $\epsilon_{i}^\theta$ are material-specific scaling and shifting parameters \textit{learned} for the corresponding layer $i$.

\begin{figure}[t!]
  \includegraphics[width=\columnwidth]{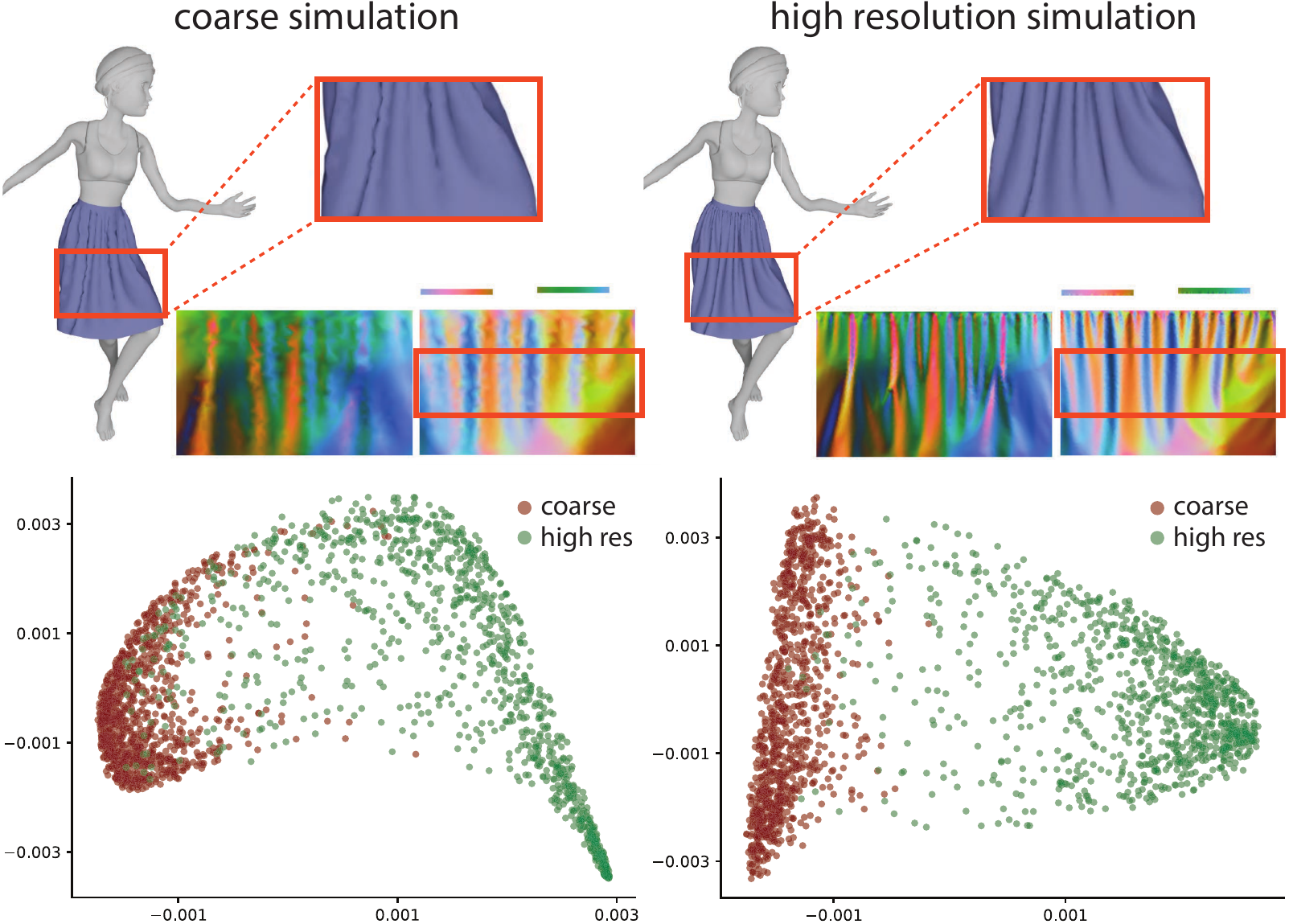}
  \caption{We pass a given set of normal map patches cropped from coarse and high resolution simulation results through VGG19. Here, we show spectral embedding in $\mathbb{R}^2$ of the feature outputs of candidate layers, and select the layers where the two sets that differ significantly to compute the style loss. We show the feature plots of two such layers. This difference is largely responsible for the drop in  visual fidelity going from high resolution to coarse simulation. }
  \label{fig:GramLayers}
\end{figure}

\paragraph{Loss Function.}  
In general, the loss function for a style transfer task is composed of two parts: (i)~content signal, which in our case penalizes the difference between $\mathbb{N}^*$ and $\mathbb{N}$ with respect to the coarse geometry features; and (ii)~style term, which in our case penalizes the perceptual difference between $\mathbb{N}^*$ and $\{\mathbb{N}_{\theta}\}$, a set of training examples that depict detailed garment geometries of material $\theta$. As indicated by Gatys et al.~\shortcite{gatys2016image}, content and style losses can be formulated using the output of a predefined set of layers, $L_C$ and $L_S$, of a  pre-trained image classification network, e.g., VGG19~\cite{simonyan2014very}. Note that the definition of content and style in our problem is not the same as in a classical image style transfer method. We performed a qualitative study to find the appropriate layers $L_C$ and $L_S$. Specifically, we pass examples of two sets of normal maps that depict coarse and fine geometry of garments of a particular material through VGG19. For candidate layers, we embed the output features in 2D space using a spectral embedding method and identify the layers where the two sets are well separated (see Figure~\ref{fig:GramLayers}).

Given $L_C$ and $L_S$, we define content loss between ${\mathbb{N}}$ and ${\mathbb{N}^*}$ as
\begin{equation*}
    L_\textit{content}({\mathbb{N}^*}|{\mathbb{N}}) := \sum_{l\in L_C} [F_l[{\mathbb{N}^*}]-F_l[{\mathbb{N}}]]^2/2,
\end{equation*}
where \(F_l\) is the VGG features of the selected layer $L_C$. 

We define the style loss based on the correlation of the VGG features across different channels by using \emph{Gram} matrices. Since the input to our network is a normal map that is generated based on a UV parameterization of the garment, it consists of background regions which do not correspond to any UV patch. In order to avoid such empty regions impacting the correlation measure, we adopt a spatial control neural style transfer method by using a UV mask $M$ that masks out the background regions. We capture style by a set of Gram Matrices defined for each $L_S$ layer of VGG:
\begin{equation*}
    G_l^M({\mathbb{N}}) := (B^M_l\circ F_l({\mathbb{N}}))^T (B^M_l\circ F_l({\mathbb{N}})),
\end{equation*}
where $B_l^M$ is the background mask for layer $l \in L_S$ downsampled accordingly. The style loss is then formulated as:
\begin{multline*}  
L_\textit{style}({\mathbb{N}^*}|{\{\mathbb{N}_{\theta}\}},M) := \\ 
\sum_{k\in\|\{\mathbb{N}_{\theta}\}\|} \sum_{l\in\{L_S\}}\frac{1}{4R_l^2}[G^M_l({\mathbb{N}^*})-G^M_l({\mathbb{N}_{\theta}^k})]^2,
\end{multline*}
where $R_l$ is the number of valid pixels in the down-sampled mask $B_l^M$.

The final loss function we use to train the network is a weighted sum of the content and style losses:
\begin{equation*}
    \text{Loss}_\Phi := w_\textit{c} L_\textit{content}({\mathbb{N}^*}|{\mathbb{N}})+w_\textit{s} L_\textit{style}({\mathbb{N}^*}|{\{\mathbb{N}_{\theta}\}},M),
\end{equation*}
where $w_\textit{c}$ and $w_\textit{s}$ denote the weight of each term respectively. In our tests, we use $w_c=1$ and $w_s=10^4$.


\paragraph{Patch-based generalization.}
Generalizing a CNN based neural network to different types of clothes with unknown UV parameterization is challenging. Particularly, similar 3D garments (e.g., a short and a long skirt) might potentially have very different parameterizations due to different sewing patterns that vary in the number of 2D patterns and their shape. In order to generalize our approach across such varying 2D parameterizations, we adopt a patch-based approach. Specifically, instead of operating with complete normal maps, we use patches cut from the normal maps as input and output of our network. While we use randomly cut patches during training, at test time we use overlapping patches sampled regularly on the image plane. We resize each input patch to ensure the absolute size of one pixel remains unchanged. The output patches that are enhanced by our network are then combined to generate the final detailed normal map $\mathbb{N}^*$ by averaging the RGB values in the overlapping regions of the output patches. 

\begin{figure}[b!]
  \includegraphics[width=\columnwidth]{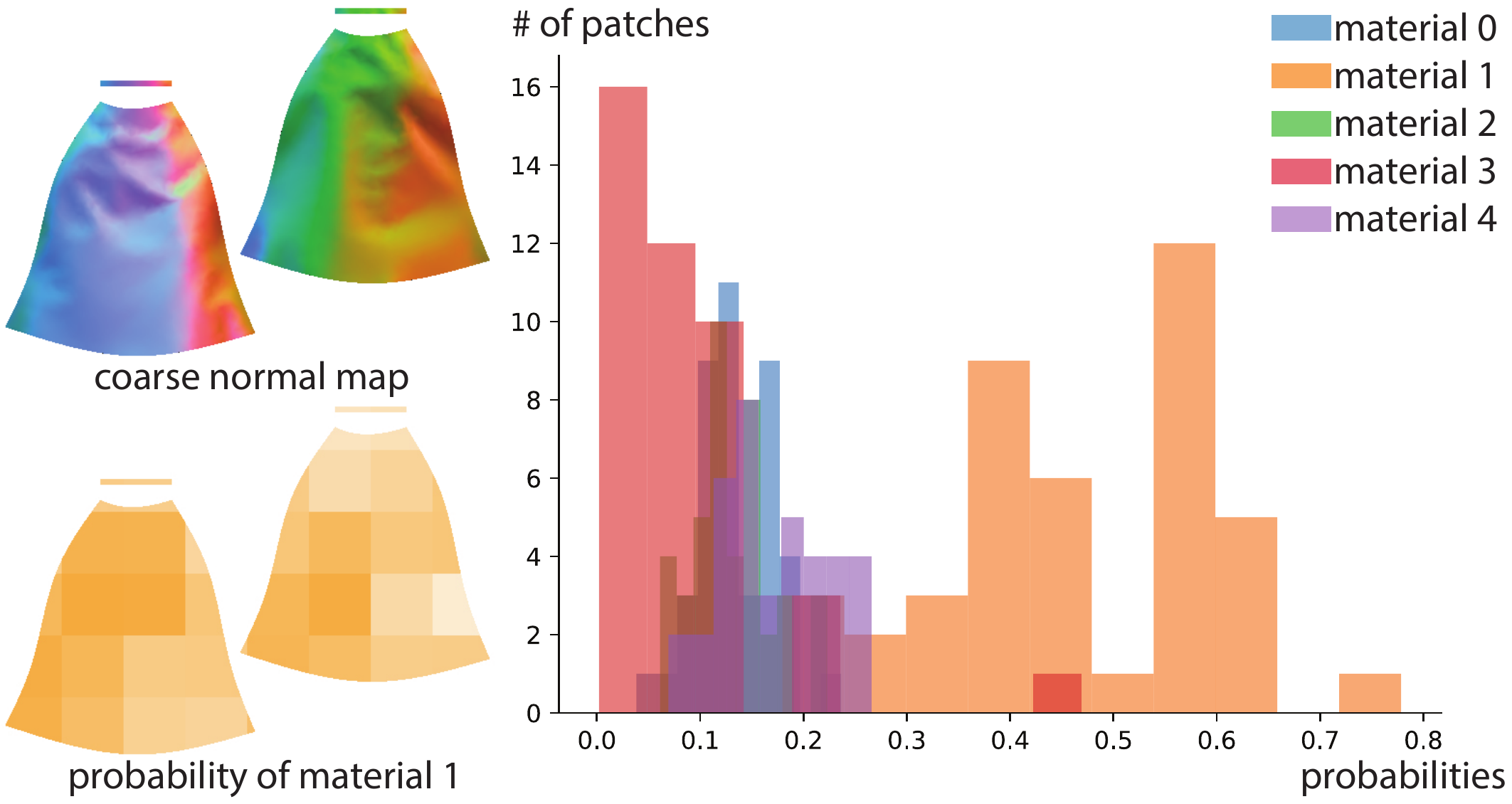}
  \caption{For an input coarse normal map, we crop 42 patches as the input of our material classification network. For each material type in our dataset, We plot the histogram of patches based on the probability of them belonging to that material. As shown, material 1 has the highest distribution among the five materials. We also color code the patches based on their probability of belonging to material 1, where darker color means a higher probability.}
  \label{fig:patchvoting}
\end{figure}

\subsection{Material Classification}
\label{subsec:garmentClassify}
 We introduce a patch-based material classification method to predict the material properties in such cases. Specifically, we train a patch-based material-classification network, $\Pi$, which is composed of $4$ fully-connected layers as shown in Figure~\ref{fig:overview}. The input to this network is the flattened feature vector of VGG calculated from a patch cropped from $\mathbb{N}$. The dimension of the last layer is the number of the materials captured in our dataset. The last layer is followed with a \textit{Sigmoid function} and a normalization layer to output the probability of the input patch belonging to each material. During training, we use \textit{Cross-Entropy} loss between the predicted probabilities and the ground truth material type. 

Humans perceive the characteristics of a material when they observe a garment in motion. Similarly, we observe that performing material classification on a single patch is not sufficiently accurate~\cite{yang2017learning}. Hence, for a given input garment sequence, we choose randomly cropped patches and run material classification on each of them sampled over time. We use a voting strategy to decide the final material type from all such individual predictions as shown in Figure~\ref{fig:patchvoting}.
Note that in application scenarios, e.g., if the coarse garment geometry is obtained by deforming a template via linear blend skinning, the material classification frees the user from having to guess material parameters. 

\subsection{3D Recovery}
\label{subsec:3dlifting}
\begin{figure*}[h]
  \includegraphics[width=\textwidth]{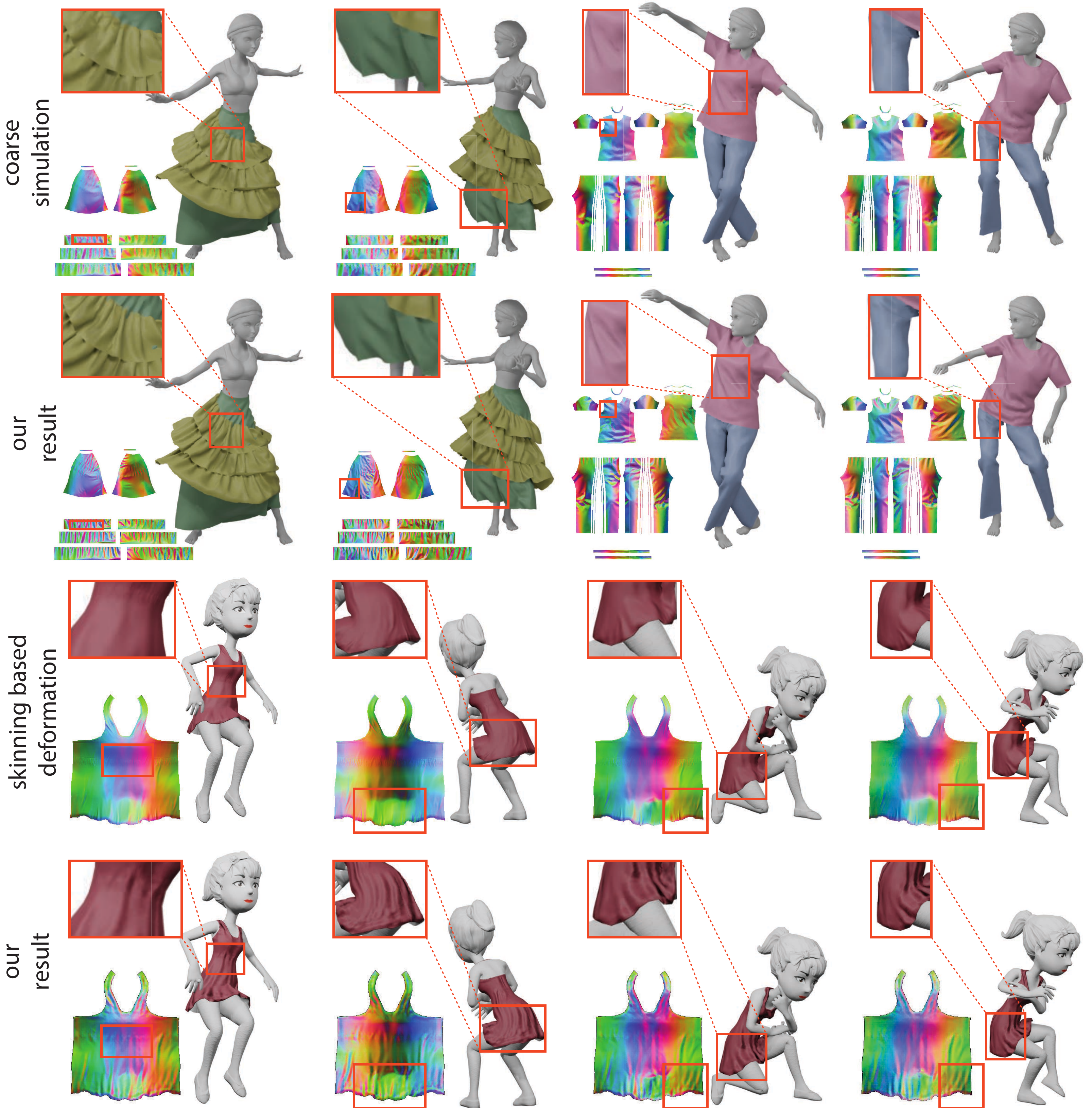}
  \caption{We evaluate our method both on coarse garment simulations and garments deformed by linear blend skinning. Our method can generalize across different garment types and motions. The materials for the simulated garments are as follows: t-shirt (knit terry), pants (wool melton), skirt (silk chamuse), skirt laces (silk chiffon). For the skinning based example, our material classification network predicts it as silk chiffon. Please see supplementary video. }
  \label{fig:result1}
\end{figure*}

\begin{figure*}[t!]
  \includegraphics[width=\textwidth]{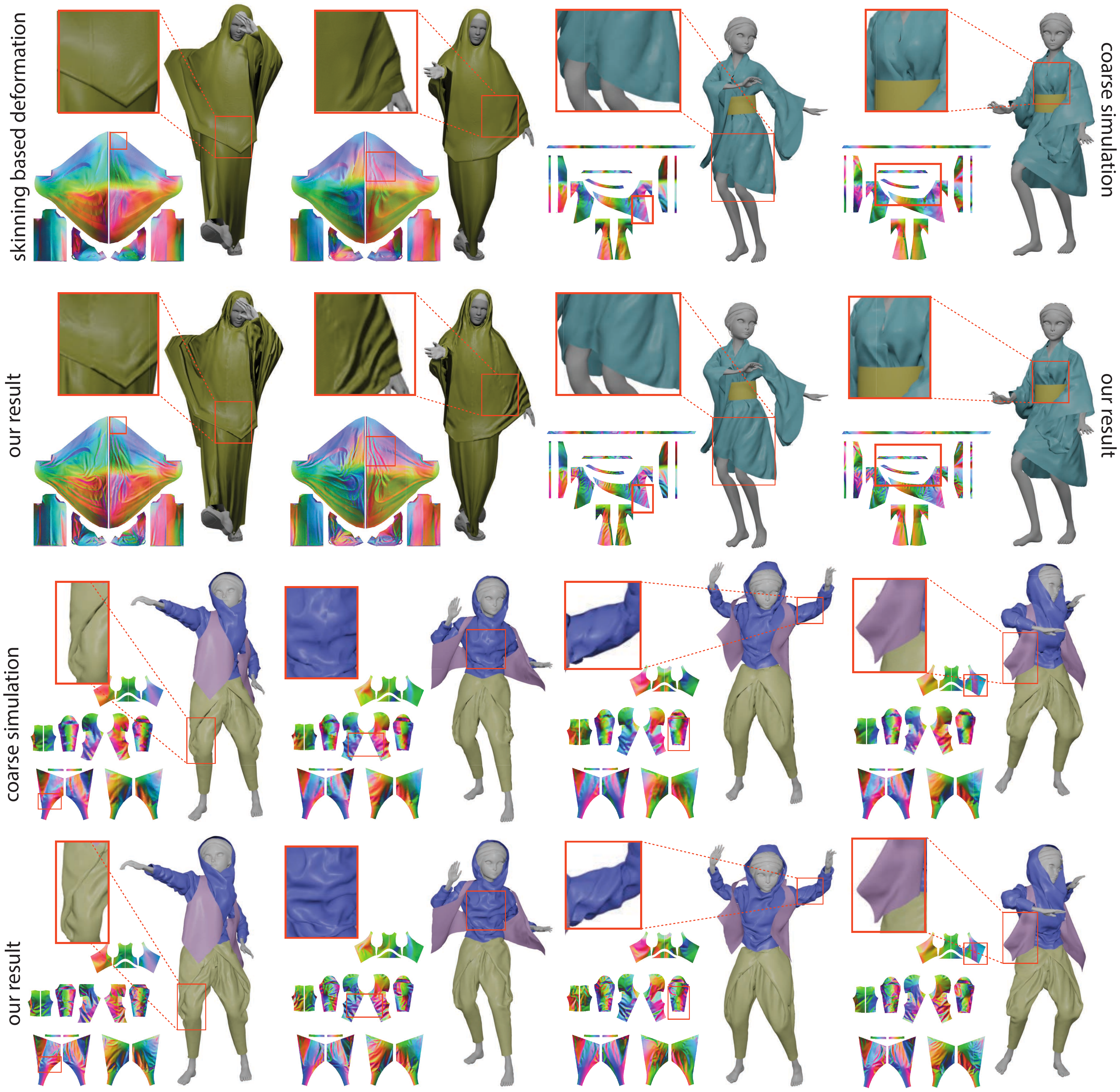}
  \caption{We evaluate our method both on coarse garment simulations and garments deformed by linear blend skinning. The materials for the simulated garments are as follows: hanfu (silk chamuse), hood (denim lightweight), pants (knit terry), vest (wool melton). For the skinning based example, our material classification network predicts it as knit terry. Please see supplementary video. }
  \label{fig:results2}
\end{figure*}

Once the enhanced normal map $\mathbb{N^*}$ is calculated, we finally lift the details into 3D. Since coarse geometry $\mathbb{S}$ is known, one common solution in the CG industry is to bake the normalmap onto the original coarse surface during rendering. The skinning based examples shown in Figure~\ref{fig:result1} and~\ref{fig:results2} are rendered using this approach. On the other hand, real 3D results are useful in many follow up applications, such as animation, texture painting, etc. We adopt a two-stage approach to recover the detailed 3D geometry for such applications. 

\paragraph{Normal map guided deformation.} We perform upsampling over the surface of the coarse geometry $\mathbb{S}$ and deform the set of vertices $P$ based on the enhanced normal map $\mathbb{N^*}$. Our method uses an iterative gradient optimization scheme to find the deformed vertex positions $p$ which minimizes the following energy:
\begin{equation*}
\argmin_{\{P\}} \sum_{p\in \{P\}}{\sum_{q\in R^1_{p}}\|n_p\cdot \frac{q-p}{\|q-p\|}\|^2} + \eta \sum_{p\in \{P\}}\|\Delta p\|^2 + \omega \sum_{p\in \{P\}}\|p_0 - p\|^2 ,
\end{equation*}

where $n_p$ is the normal of vertex $p$ as denoted on $\mathbb{N^*}$, $R^1_{p}$ is the one-ring neighborhood of $p$, \(\Delta\) is the Laplacian operator, and $p_0$ is the initial position of $p$ obtained by upsampling the coarse geometry, and  \(\eta\) and \(\omega\) are the weights of the last two regularization terms. While the Laplacian term acts as a smoothness term to avoid sharp deformations, the last term penalizes the final shape from deviating from the original geometry (e.g., by shifting or scaling). 

\paragraph{Avoiding interpenetration.} The above deformation approach may cause interpenetrations between the garment and the underlying body surface. Hence, we adopt a post-processing step to resolve such penetrations as introduced by Guan et al.~\shortcite{guan2012drape}. Specifically, we first find the set of garment vertices $P_C \in P$, which are inside the body mesh. We project each such vertex $p \in P_C$ to the nearest location, $p^*$, on the body surface. We update the vertex positions $p$ such that the following objective function is minimized:
\begin{equation*}
\argmin_{\{p\}} \sum_{p\in \{P_C\}}\|p^*-p\|^2 + \phi \sum_{p\in \{P\}}\|\triangledown p\|^2.
\end{equation*}
In our experiments, we use $\eta=5\times10^3$, $\omega=8$, and $\phi=5\times10^3$. The objective functions are optimized by standard gradient descent algorithm.

\subsection{Data Generation}

\begin{figure}[t!]
  \includegraphics[width=\columnwidth]{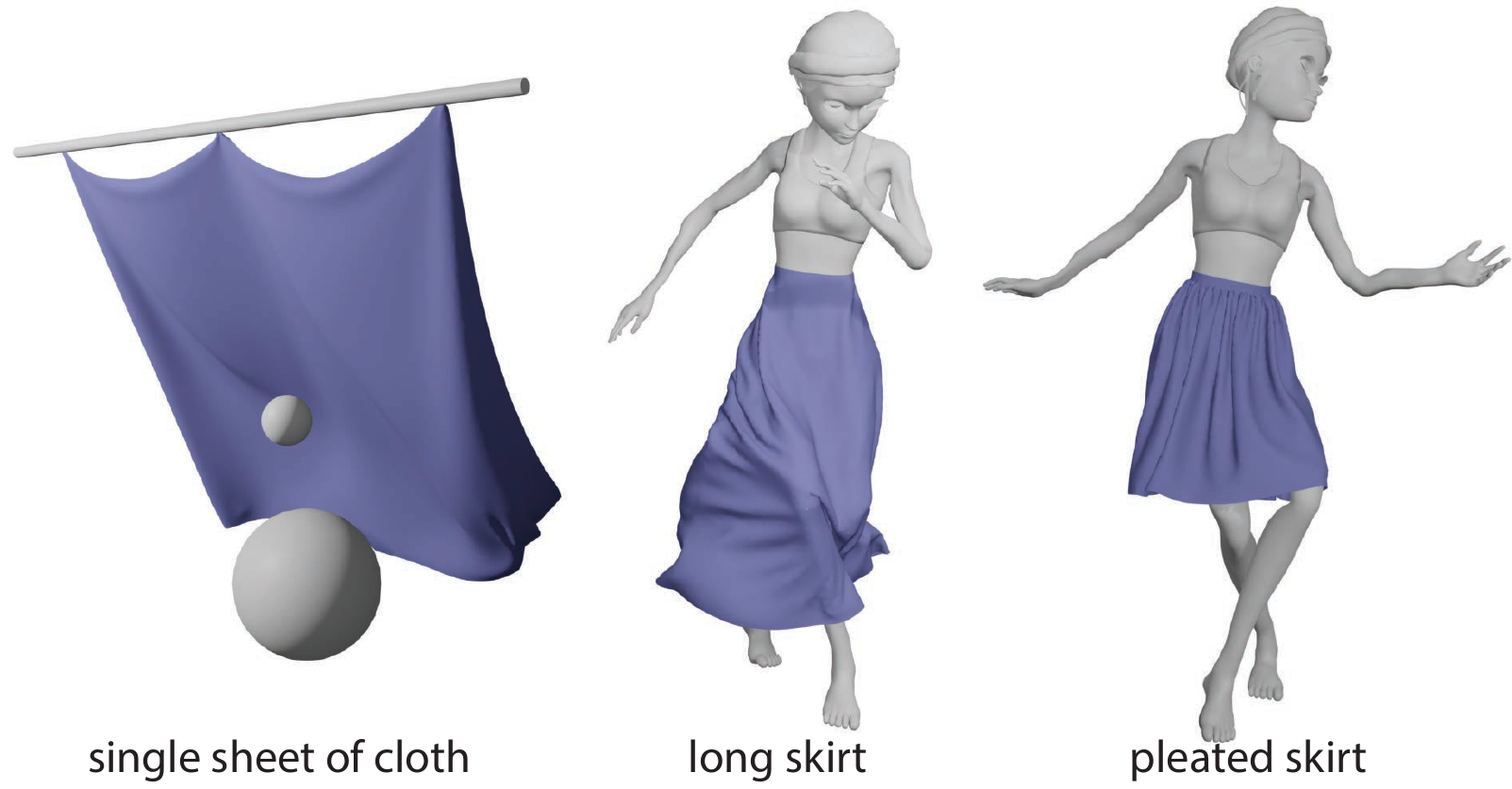}
  \caption{Our base training data is simulated from a single sheet of cloth interacting with two balls and blowing wind. We also simulate a long skirt and a pleated skirt under different body motions. For each training example, we simulate both a coarse and a fine resolution mesh and obtain the corresponding normal map pairs. }
  \label{fig:trainingData}
\end{figure}

Our goal is to generalize our method to a large variety of garment types and motion sequences. To ensure such a generalization capability, we generate our training data by performing physically-based simulation over 3 different garment-motion pairs (see Figure~\ref{fig:trainingData}). These are:
\begin{enumerate}
    \item a piece of hanging sheet interacting with 2 balls and blowing wind ($901$ frames at $30$ fps);
    \item a long skirt over a virtual character performing Samba dance ($988$ frames at $30$ fps);
    \item a pleated skirt over a virtual character performing modern dance ($850$ frames at $30$ fps).
\end{enumerate}   
For the last two cases, the underlying virtual character is created by Adobe Fuse\footnote{\url{https://www.adobe.com/products/fuse.html}} and the motion sequences are obtained from Mixamo\footnote{\url{https://www.mixamo.com/}}. We use Marvellous Designer\footnote{\url{https://www.marvelousdesigner.com}} to drape the garments over the 3D character body and perform the physically-based simulation. We simulate each garment with different material samples (silk chamuse, denim lightweight, knit terry, wool melton, silk chiffon) and in two resolution settings. To obtain the coarse geometry, we run the simulation with a particle distance of $30mm$. This results in resulting meshes of $4440,2287,1756$ vertices for the aforementioned 3 types of garments respectively. To generate the fine scale geometry, we run the simulation with a particle distance of $10mm$ which results in meshes with $39330,18950,14433$ vertices respectively.
The normal map of the 3D garment surface is then calculated. For each example, we generate the corresponding normal map using a given uv parameterization for each garment type. 
We scale the normal maps of different examples to ensure a single pixel corresponds to the same unit in 3D and randomly cut patches from the normal maps and use each pair of non-empty coarse and high resolution patches as input and target of our network. Finally, we also augment our dataset by rotating and reflecting the patches. 

We observe that, sometimes due to instabilities in the physical simulation, simulation results with different particle distances may reveal geometric variations, e.g., in the position and the shape of the wrinkles. This results in input and output patch pairs that are not well aligned. To avoid this problem, we instead downsample the normal maps of the high resolution mesh to generate the coarse normal maps that will be provided as input to our network. During testing, the inputs to our network are obtained from the coarse simulation results. Our experiments demonstrate that training with the downsampled normal maps while predicting with the coarse normal maps as input at test is a successful strategy.

\paragraph{Network architecture.}  
Our network broadly consists of an encoder and a decoder. There are 4 layers in the encoder to project the input patch image to a compact latent space, by down-sampling from \(128 \times 128\) to \(8 \times 8\). In each layer, there are blocks in the order of 2D convolution, Relu, and maxpool. The followed decoder is used to up-sample the patch tensor from \(8 \times 8\) to the original size, mainly using 2D trans-convolutions, and a 2D convolution in the last layer. Because different materials share one network, we adopt CIN layer to scale and shift the latent code to the specified material related space, after every 2D trans-convolution. Finally,  we  use skip connection, as we regard our problem as constructing the residual for the fine image based on the coarse one.  In our experiments, we use the Adam  optimizer, with a learning rate of $1e-4$ and parameters $\beta_1=\beta_2=0.9$. Project code and data will be released. 

\begin{figure*}[t!]
  \includegraphics[width=\textwidth]{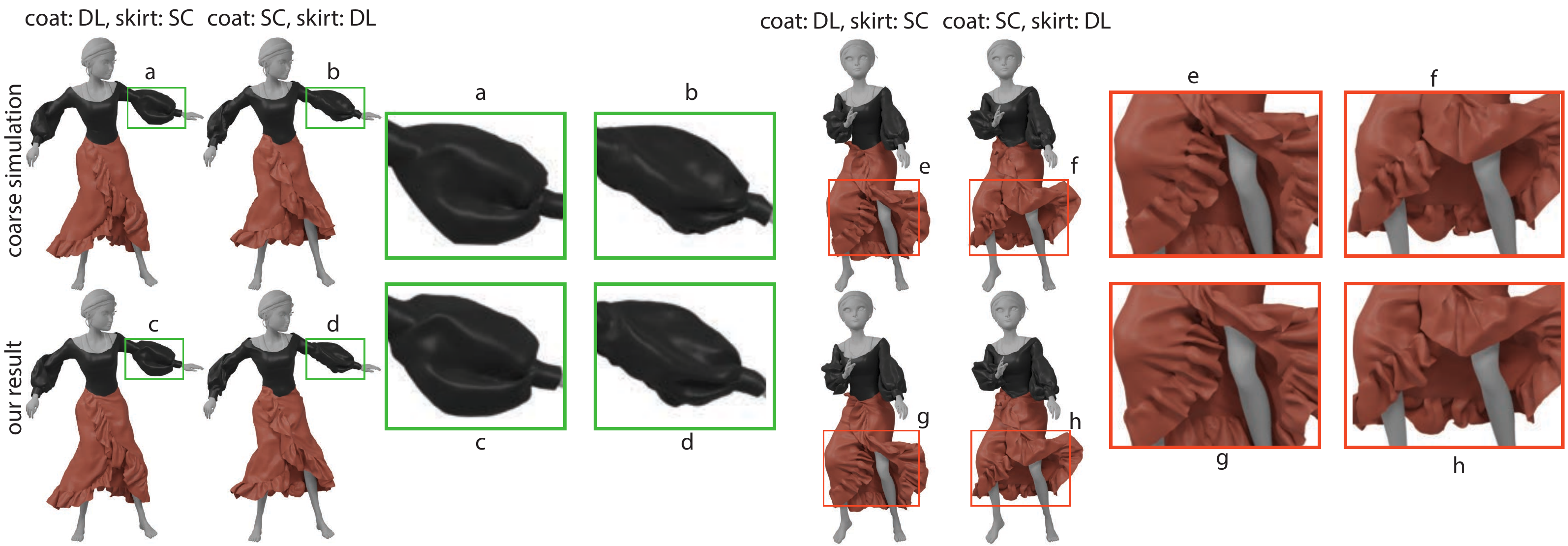}
  \caption{We demonstrate how our method works on garments made of different materials and undergoing the same motion. Note how the coarse inputs vary with material change and the synthesized details that align well with the input. The two different materials used are denim light (DL) and silk chiffon (SC).}
  \label{fig:material}
\end{figure*}

\section{Results}
\label{sec:evaluation}
We evaluate our method on the product space of a variety of garments undergoing various body motions. We present qualitative and quantitative evaluation, comparison with baseline methods, and also ablation study. 

\subsection{Dataset}
As input, we test on coarse garment geometries obtained by either running a physically-based simulation on a low resolution mesh of the garment, as well as garments deformed via a skeleton rig using \textit{linear blend skinning} (LBS). Similar to the dataset generation, we use motion sequences from Mixamo and utilize Marvelous Designer as our physical simulator. For skinning-based examples, we test our method on a character from Mixamo, as well as a character from the Microsoft Rocketbox dataset~\footnote{https://github.com/microsoft/Microsoft-Rocketbox}. We show qualitative examples in Figures~\ref{fig:result1} and~\ref{fig:results2}.

\subsection{Evaluation}

\paragraph{Qualitative evaluation}
The examples demonstrate the generalization capability of our method to \textit{unseen garments and motions} during training time. Even for garment types that are significantly different from our training data, e.g., the multi-layer skirt in Figures~\ref{fig:teaser}, \ref{fig:result1}, and the hijab in Figure~\ref{fig:results2}, our network synthesizes plausible geometric details an order of magnitude faster than running a physically based simulation with a high resolution mesh (see Figure~\ref{fig:motivation}). When evaluating our method on skinning-based rigged characters, since no garment material information is provided, we use our material classification method to first predict plausible material parameters based on the input normal maps. We then utilize this material information to synthesize vivid wrinkles that enhances the input garment geometry. Empirically, even when material specification was available, we found training end-to-end with the material classification network along with CIN resulted in faster convergence~\cite{huang2017arbitrary}.

Material properties of a garment plays an important role in the resulting folds and wrinkles. Our method captures such deformations by taking the material parameters as an additional input. In Figure~\ref{fig:material}, we show examples of garments made from different materials undergoing the same body motion. Our method is able to generate plausible details in all cases that well respect the input coarse geometries.

\paragraph{Quantitative evaluation}
In order to evaluate our approach quantitatively, when ground truth data is available, we define an \emph{improvement score} given an input patch, ($\mathbb{N}$), an output patch, ($\mathbb{N^*}$), and the corresponding ground truth patch, ($\hat{\mathbb{N}}$), as follows:
\begin{equation*}
    \text{improvement score} :=
    100 \left|
    \frac{L_\textit{style}({\mathbb{N}})-L_\textit{style}({\mathbb{N^*}})}{L_\textit{style}({\mathbb{N}})-L_\textit{style}({\hat{\mathbb{N}}})} \right|.
\end{equation*}

\begin{table}[b!]
\caption{We report the mean and standard deviation of improvement scores on three dresses made of two different materials and undergoing two different motion sequences. While Dress A and Motion 1 was seen during training, the remaining dresses and the motion sequence are unseen. Both motions are sequences are composed of 200 frames. }
\begin{tabular}{|c|c|c|c|c|}
\multicolumn{1}{c}{} & \multicolumn{1}{c}{} & \multicolumn{1}{c}{\includegraphics[width=0.15\columnwidth]{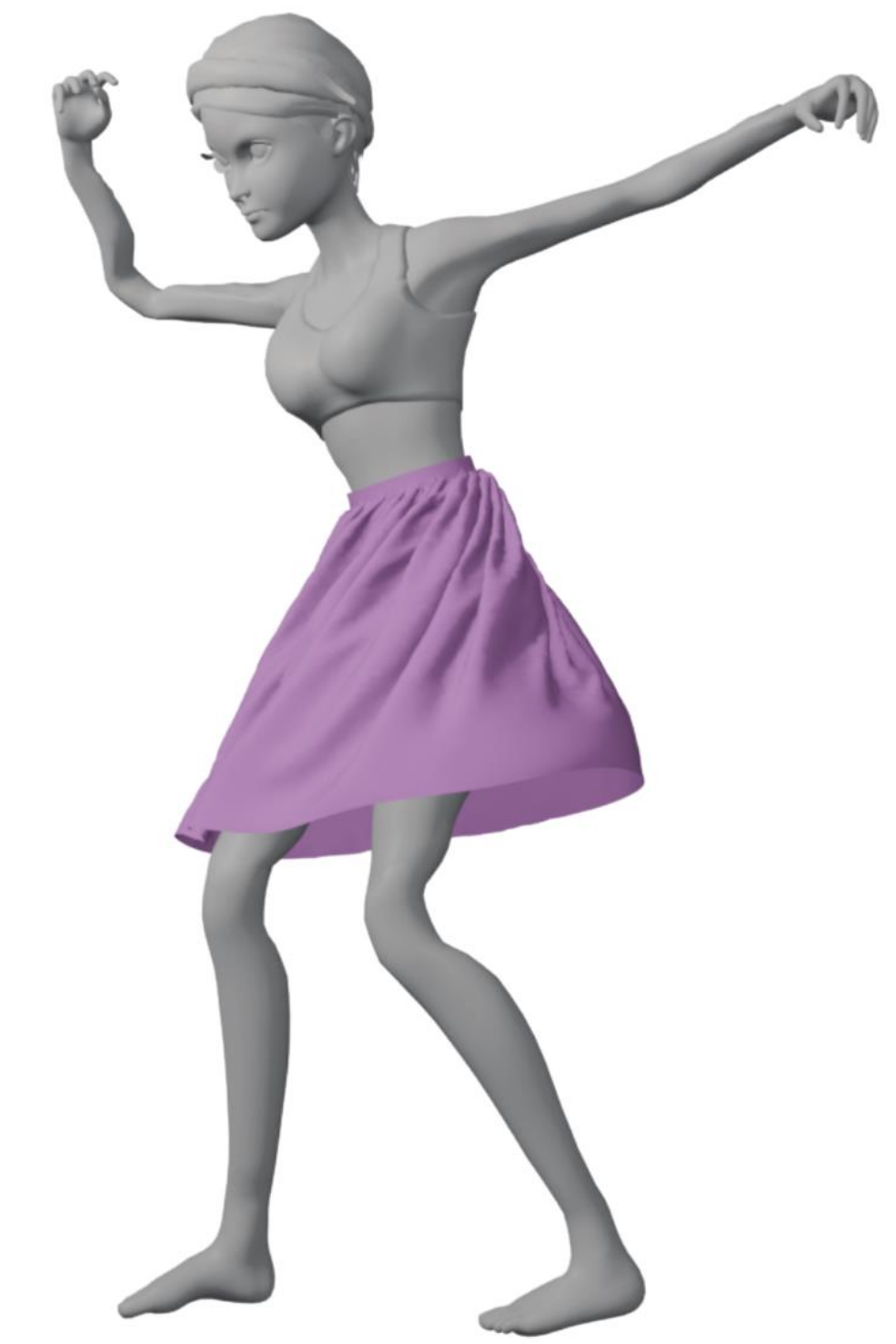}} & \multicolumn{1}{c}{\includegraphics[width=0.15\columnwidth]{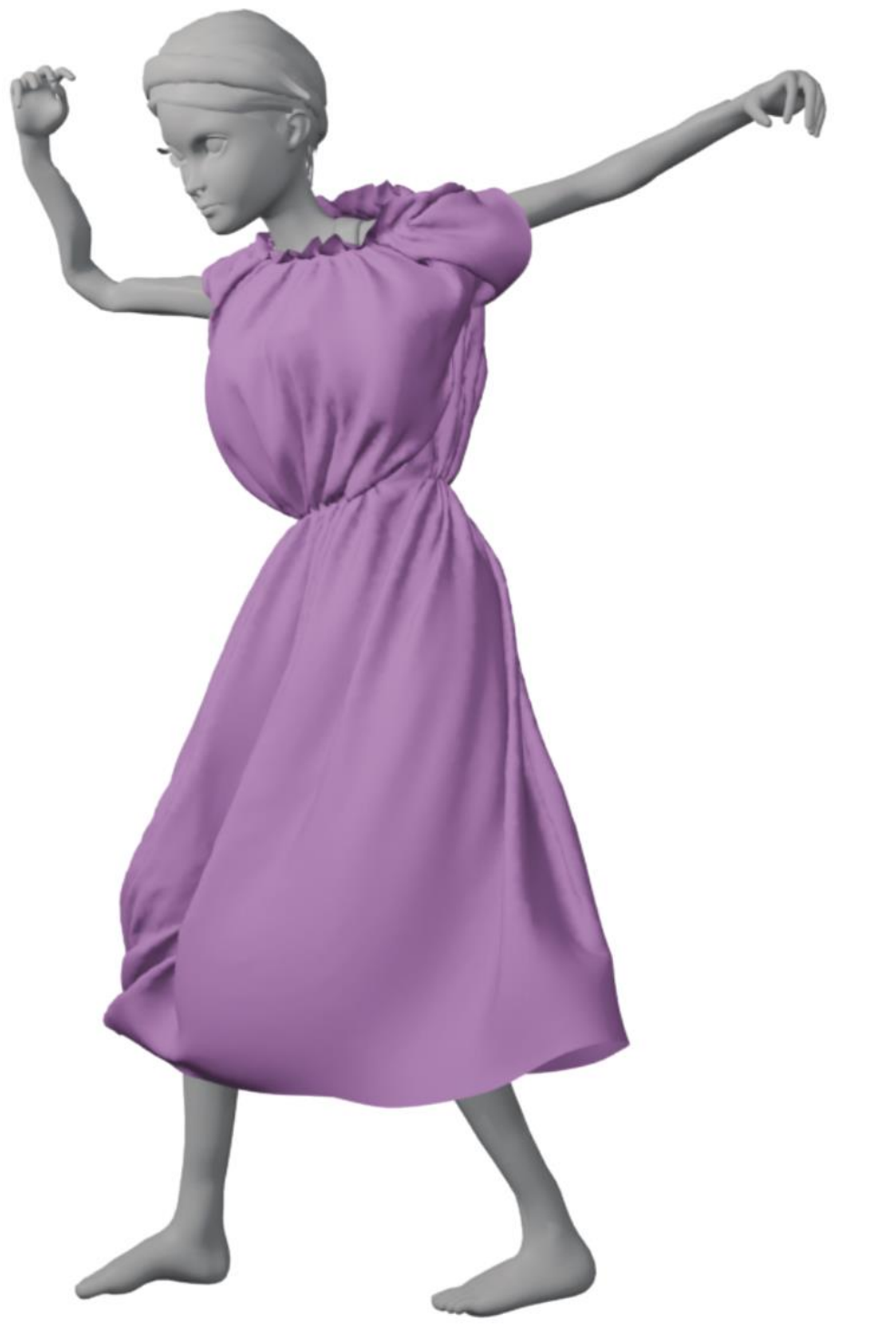}} & \multicolumn{1}{c}{\includegraphics[width=0.15\columnwidth]{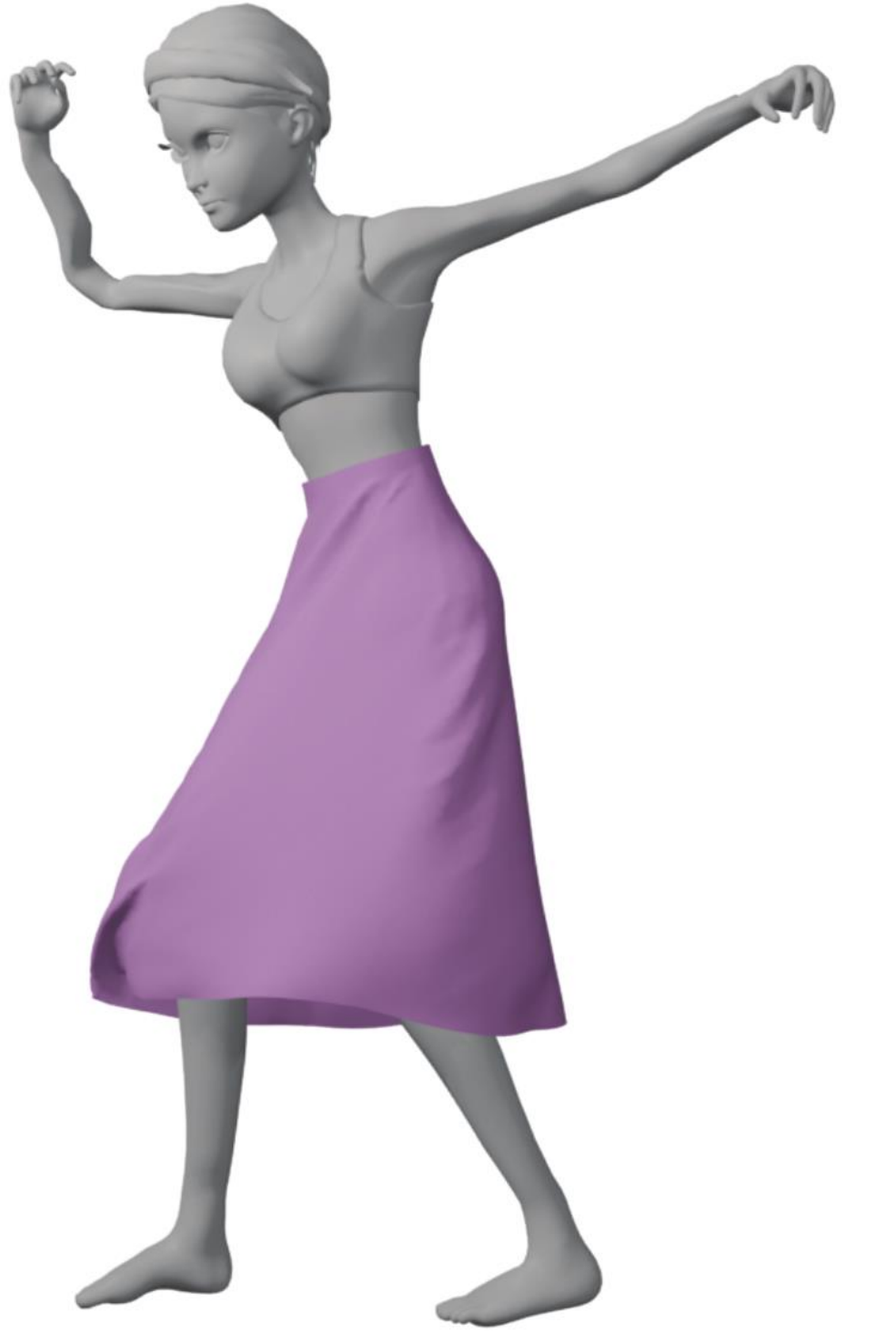}} \\
\hline
   & \diagbox[width=5em]{Mean}{Std} & Dress A & Dress B & Dress C\\ 
\hline
\multirow{2}{*}{\STAB{\rotatebox[origin=c]{90}{\makecell{Silk\\ chamuse}}}} & Motion 1 & \diagbox[width=5em]{98.41}{0.87} & \diagbox[width=5em]{96.22}{1.01} & \diagbox[width=5em]{91.87}{3.02} \\ \cline{2-5}
& Motion 2 & \diagbox[width=5em]{97.55}{1.53} & \diagbox[width=5em]{95.39}{0.92} &\diagbox[width=5em]{91.25}{4.56} \\ \hline
\multirow{2}{*}{\STAB{\rotatebox[origin=c]{90}{\makecell{Denim \\light}}}} & Motion 1 & \diagbox[width=5em]{98.35}{0.57} & \diagbox[width=5em]{95.28}{0.92} & \diagbox[width=5em]{92.45}{2.03}\\ \cline{2-5}
& Motion 2 & \diagbox[width=5em]{97.39}{1.08} & \diagbox[width=5em]{95.25}{1.42} & \diagbox[width=5em]{92.44}{3.25} \\\hline
\end{tabular}
\label{table:quantitative}
\end{table}

A higher improvement score, 100 being perfect, indicates that the output patch is closer to the ground truth in terms of the styles of the wrinkles. Table~\ref{table:quantitative} reports the mean and standard deviation of the improvement score for three different garment types made from two different materials and undergoing two different motion types. While one garment and motion type has been seen during training (dress A, motion 1), the remaining garment types and the motion sequence are unseen. Not surprisingly, the method performs well~($\sim98\%$) on seen garment types and motions. The main result is the generalization to unseen data and achieves high improvement scores regardless with only slight degradation in terms of performance ($\sim91-96\%$).

When ground truth is not available, we evaluate the performance from a statistics perspective. Given a sequence of animated garments, we randomly crop patches from each frame and generate the corresponding samples of Gram matrices of style features. We collect such samples from three sources: (a)~patches from the coarse input; (b)~patches generated by our method; and (c)~patches from high resolution simulation (not for corresponding frames). We embed the patches using spectral embedding from $\mathbb{R}^{262144} \rightarrow \mathbb{R}^2$, and measure the distances in the distribution of the resultant point sets (denotes as $P_a$, $P_b$, and $P_c$, respectively) using Chamfer distance~(CD) as $C_1 := \text{CD}(P_a,P_c)$ and $C_2 := \text{CD}(P_b,P_c)$, and improvement score defined as $
\text{DR} := 100 {|C_1 - C_2|}/{C_1}$ in Table~\ref{table:quantitative2}. As shown in the table, the Chamfer distance between our result and the ground truth is significantly lower than the distance between the coarse input and the ground truth. This indicates that our method not only enhances the details of individual frames, but also improves overall plausibility for a motion sequence.

\begin{table}[h]
\caption{ We report the Chamfer distance between (i)~the coarse input and the ground truth and (ii)~our result and the ground truth. We calculate a matching score \{$\text{DR} := 100 |C_1 - C_2|/C_1$\} to show the relative improvement. Please refer to the text for details.  Note that Dress A and Motion 1 are seen during training while the remaining dresses and the motion sequence are unseen. Dress C is tight, as indicated by the low Chamfer distance to start, and therefore leaves little scope to improve in terms of detail enhancement.}
\begin{tabular}{|c|c|c|c|c|}
\hline
& \diagbox[width=6em]{$C_1$}{$C_2$}{DR} & Dress A & Dress B & Dress C\\ \hline
\multirow{2}{*}{\STAB{\rotatebox[origin=c]{90}{\makecell{Silk\\ chamuse}}}} & Motion 1 & \diagbox[width=6em]{1.77}{0.09}{94.81} & \diagbox[width=6em]{0.36}{0.09}{74.32} & \diagbox[width=6em]{0.15}{0.12}{14.95} \\ \cline{2-5}
& Motion 2 & \diagbox[width=6em]{1.39}{0.11}{92.25} & \diagbox[width=6em]{1.09}{0.12}{89.20} &\diagbox[width=6em]{0.11}{0.10}{10.56} \\ \hline
\multirow{2}{*}{\STAB{\rotatebox[origin=c]{90}{\makecell{Denim \\light}}}} & Motion 1 & \diagbox[width=6em]{1.31}{0.01}{92.54} & \diagbox[width=6em]{0.89}{0.13}{85.86} & \diagbox[width=6em]{0.13}{0.12}{6.36}\\ \cline{2-5}
& Motion 2 & \diagbox[width=6em]{1.50}{0.12}{92.33} & \diagbox[width=6em]{0.93}{0.11}{88.23} & \diagbox[width=6em]{0.10}{0.09}{7.70} \\\hline
\end{tabular}
\label{table:quantitative2}
\end{table}

\paragraph{Continuity in time}
While we do use time information in the material classification, the main detail enhancement step works at frame level. In earlier versions of our network, we experimented with time-dependent networks, with GRUs and LSTMs. While the results were better on training sets, we found the networks did not generalize across unseen data. In contrast, the proposed per-frame processing still leverages continuity data in the coarse geometry sequence and produces time-continuous details (please see supplementary video), with superior generalization behavior. We quantitatively evaluate our performance over continuous body motion, see  Figure~\ref{fig:trainingData} (right). 

\subsection{Ablation Study}
\begin{figure}[t!]
  \includegraphics[width=\columnwidth]{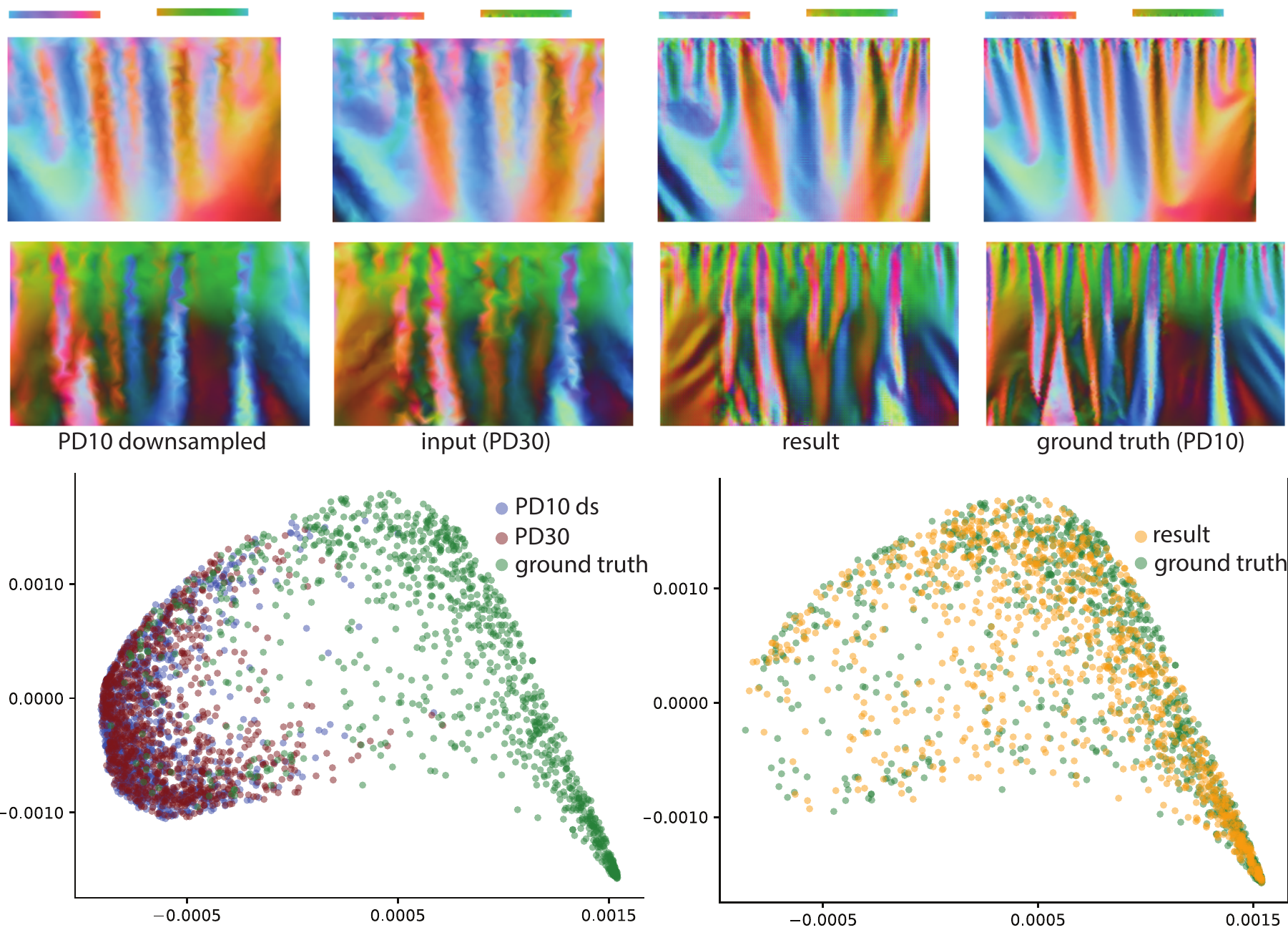}
  \caption{Given an input normal map obtained by running a coarse simulation with a particle distance of 30 mm (PD30), our network produces a result that is close to the ground truth (high resolution simulation run with a particle distance of 10 mm) in terms of Gram matrix distribution of the output features of VGG19 layers that we use as style layers. We note that when training our network, as input we use downsampled high resolution normal maps (PD10 ds) which have similar feature distribution as the coarse simulation normal maps (PD30) used during test time.}
  \label{fig:vgg_tse_visualization_result}
\end{figure}

In order to ensure coarse and high resolution patches are aligned during training, we generate the input to our network by downsampling the high resolution normal maps. We validate this choice, by visualizing the 2D embedding of the output features of some of the layers of VGG19 that we choose as style layers in Figure~\ref{fig:vgg_tse_visualization_result}. As shown in the figure, the distribution of the features of the downsampled high resolution normal maps are close to the actual coarse normal maps. Further, given the downsampled normal maps as input, our method can successfully generate results that share similar distribution as the ground truth.

\begin{figure}[b!]
  \includegraphics[width=\columnwidth]{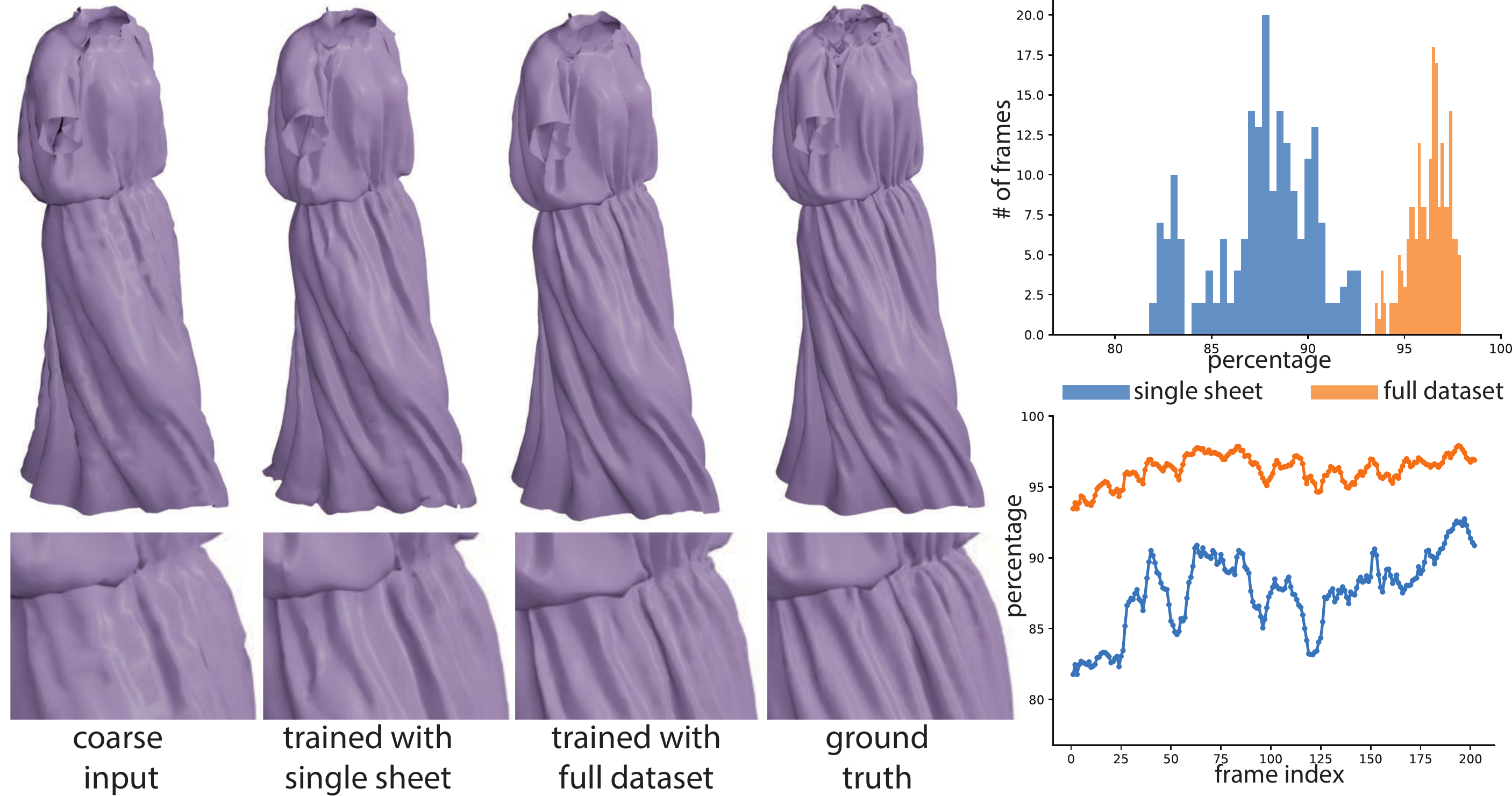}
  \caption{We show results of our method when trained with (i)~data generated from a single sheet only and (ii)~all the three datasets~(see Figure~\protect\ref{fig:trainingData}).}
  \label{fig:ablation}
\end{figure}

\begin{figure*}[t!]
  \includegraphics[width=\textwidth]{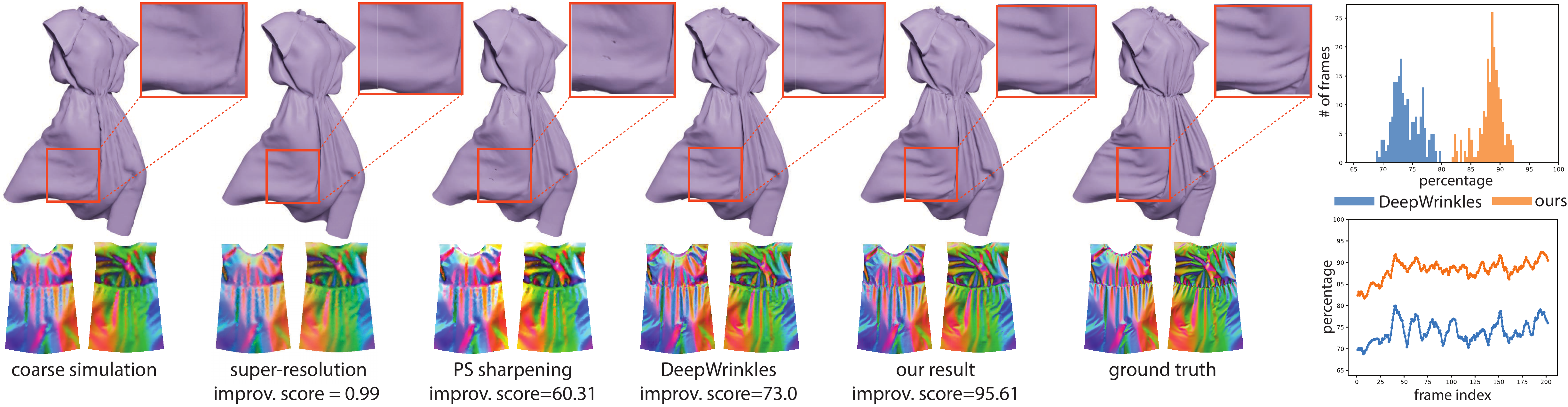}
  \caption{Given a coarse garment geometry, we compare our results with image based methods, i.e., sharpening feature in Photoshop and image super-resolution~\cite{Wang_2018_CVPR_Workshops}, as well as our implementation of the DeepWrinkles~\cite{lahner2018deepwrinkles} method. Our results are closest to the target ground truth visually. We also outperform the alternative approaches in terms of the improvement score.}
  \label{fig:comparisons}
\end{figure*}

We also evaluate the performance of our method trained with different portions of our dataset. As shown in Figure~\ref{fig:ablation}, even when trained with a single sheet of cloth, our method generalizes reasonably well to different garment types. The quality of the results improve as the complete dataset is utilized.

\subsection{Comparison}
In Figure~\ref{fig:comparisons}, we compare our method to other alternative approaches. Specifically, we compare to two image-based enhancement methods, namely \emph{image sharpening} feature in Adobe Photoshop~\cite{photoshop} and a state-of-the-art image super-resolution method~\cite{Wang_2018_CVPR_Workshops}. 
We observe that image-based approaches cannot really hallucinate wrinkle details and are not really suitable for our problem. 

We further compare our method to the recent neural network-based approach of DeepWrinkles~\cite{lahner2018deepwrinkles}. We implement their method and train both their generative model and our network with a portion of our dataset, specifically with training data obtained from the long skirt. While DeepWrinkles, which uses a patch-based GAN setup, achieves similar performance during the training stage, it falls short when generalizing to unseen cases. We also report the improvement score across the whole motion sequence of each method demonstrating the superiority of our approach quantitatively.

\begin{figure}[h]
  \includegraphics[width=\columnwidth]{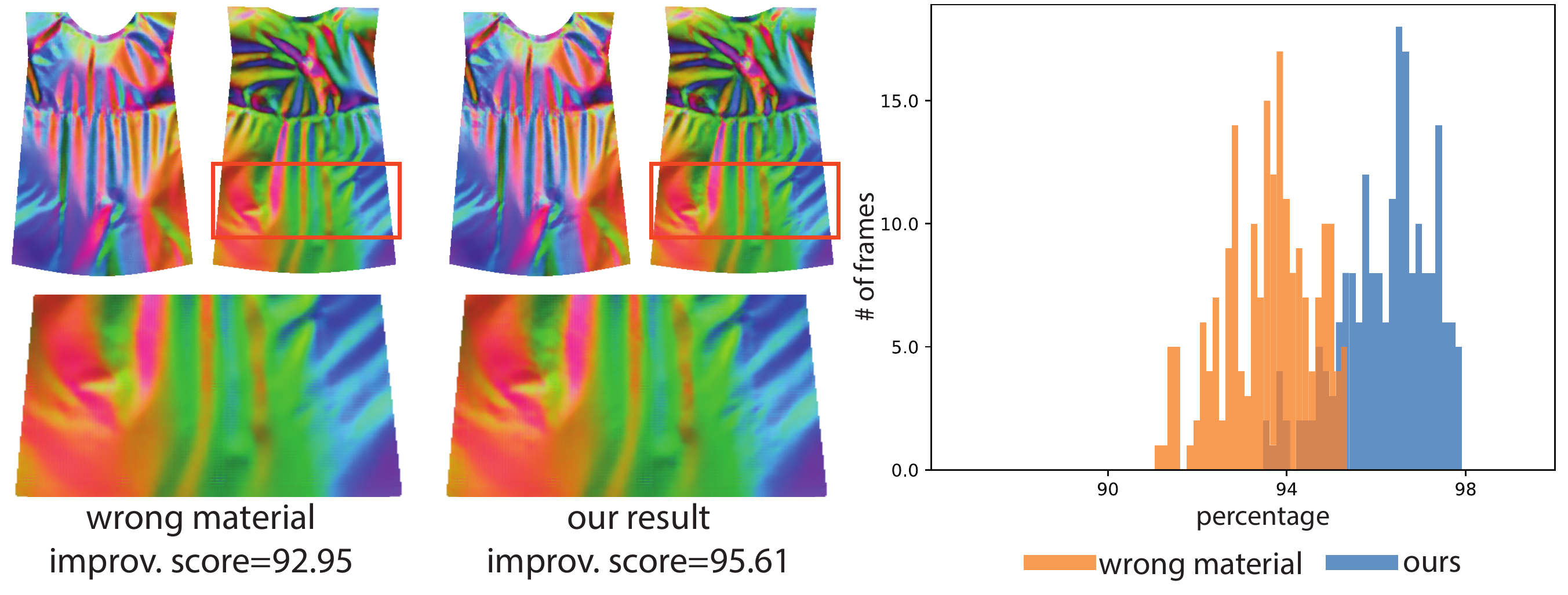}
  \caption{We enhance the same coarse normal map respectively with right (silk chamuse) and wrong (wool melton) materials. The method cannot recover subtle wrinkle details of the silk material when wrongly assigned a different material.}
  \label{fig:wrongMat}
\end{figure}

Finally, given the same coarse normal map as input, we also provide the result of our method when run with incorrect material parameters and report the respective improvement scores in Figure~\ref{fig:wrongMat}. Use of wrong material parameters results in less plausible results with around $5\%$ drop in the improvement score.

\section{Conclusions and Future Work}
We have presented a deep learning based approach to synthesize plausible wrinkle details on coarse garment geometries. Our approach draws inspiration from the recent success of image artistic style methods and casts the problem as detail transfer across the normal maps of the garments. We have shown that our method generalizes across different garment types, sewing patterns, and motion sequences. We have also trained and tested our method with different material parameters under a universal framework. We have demonstrated results on inputs obtained from running physical simulation on low-resolution garment geometries as well as garments deformed by skinning. 

\paragraph{Limitations and Future work.}
While our method shows impressive generalization capability, it has several limitations that can be addressed in future work. First of all, in order to generalize the detail enhancement network across different materials, we provide the material parameters as input. Hence, at run time, the network can be tested only on material types that have been seen during training. Generalizing our method to unseen materials, e.g., via learning an implicit material basis, is an interesting future direction. The current bottleneck is getting diverse simulations by sampling a variety of realistic materials due to challenges in getting robust and realistic simulations, in reasonable time, using commercial simulation frameworks. 
In our present implementation, our method is trained with regularly cropped 2D patches on the normal maps. Cropping geodesic patches over 3D surface instead to minimize averaging errors across overlapping regions and integrating the post-processing step into the network is an exciting direction. Furthermore, we assume that the coarse and high-resolution garments share the same set of garment pieces. In a possible garment design scenario, the designer might add some accessories to an existing garment, e.g., adding a belt or a layer to a skirt. It would be interesting to hallucinate the wrinkle details after such edits using the coarse simulation of the base garment only.

\bibliographystyle{ACM-Reference-Format}
\bibliography{main}
\end{document}